\documentclass[a4paper,preprintnumbers,floatfix,superscriptaddress,pra,twocolumn,showpacs,notitlepage,longbibliography]{revtex4-1}
\usepackage[utf8]{inputenc}
\usepackage[T1]{fontenc}
\usepackage[sc,osf]{mathpazo}
\usepackage{amsmath}
\usepackage{amsthm}
\usepackage{bbm}
\usepackage{comment} 
\usepackage{marvosym}
\usepackage{graphicx}
\usepackage{xcolor}
\usepackage{appendix}
\usepackage{hyperref}

\newcommand{\ket}[1]{| #1 \rangle}
\newcommand{\bra}[1]{\langle #1 |}
\newcommand{\kb}[2]{| #1 \rangle\hspace{-2pt}\langle #2 |}

\newcommand{\beq}{\begin{eqnarray}}
\newcommand{\eeq}{\end{eqnarray}}

\newcommand{\mean}[1]{\ensuremath{\langle{#1}\rangle}}

\newtheorem{theorem}{Theorem}

\newtheorem{definition}{Definition}
\newtheorem{result}[theorem]{Result}

\DeclareMathOperator{\tr}{\mathrm{tr}}
\DeclareMathOperator{\Tr}{\mathrm{tr}}

\newcommand{\avg}[1]{\mean{#1}}
\renewcommand{\openone}{\mathbbm{1}}

\def\1{\mathbf{1}}
\def\0{\mathbf{0}}

\DeclareMathOperator{\arccot}{arccot}

\begin{document}
\title{Quantifying causal influences in the presence of a quantum common cause}
\author{Mariami Gachechiladze}
\thanks{These authors contributed equally to this work.\\ e-mail: mgachech@uni-koeln.de,\; nikolai.miklin@ug.edu.pl}\affiliation{
Institute for Theoretical Physics, University of Cologne, Germany}
\author{Nikolai Miklin}
\thanks{These authors contributed equally to this work.\\ e-mail: mgachech@uni-koeln.de,\; nikolai.miklin@ug.edu.pl}\affiliation{International Centre for Theory of Quantum Technologies (ICTQT), University of Gdansk, 80-308 Gda\'nsk, Poland}
\author{Rafael Chaves}
\affiliation{International Institute of Physics, Federal University of Rio Grande do Norte, 59070-405 Natal, Brazil}
\affiliation{School of Science and Technology, Federal University of Rio Grande do Norte,59078-970 Natal, Brazil}

\date{\today}
\begin{abstract}
Quantum mechanics challenges our intuition on the cause-effect relations in nature. Some fundamental concepts, including Reichenbach's common cause principle or the notion of local realism, have to be reconsidered. Traditionally, this is witnessed by the violation of a Bell inequality. But are Bell inequalities the only signature of the incompatibility between quantum correlations and causality theory? Motivated by this question we introduce a general framework able to estimate  causal influences between two variables, without the need of interventions and irrespectively of the classical, quantum, or even post-quantum nature of a common cause. In particular, by considering the simplest instrumental scenario -for which violation of Bell inequalities is not possible- we show that every pure bipartite entangled state violates the classical bounds on causal influence, thus answering in negative to the posed question and opening a new venue to explore the role of causality within quantum theory.
\end{abstract}

\maketitle
\emph{Introduction--} Estimating relations of cause and effect are central and yet one of the most challenging goals of science. Since long it has been realized that correlations do not imply causation. The reason is that any correlation observed between two or more random variables can, at least in the classical regime, be explained by a potentially unobserved common cause. Understanding under which conditions such confounding factors can be controlled, such that empirical data can be turned into a causal hypothesis, has found a firm theoretical basis with the establishment of the mathematical theory of causality \cite{pearl2009causality,spirtes2000causation}. Concepts like interventions, randomized controlled experiments and instrumental variables are nowadays basic worktools in the estimation of causal influences in a variety of fields \cite{pearl2009causal,glymour2001mind,morgan2015counterfactuals,shipley2016cause,peters2017elements}.

Despite its success, all such ideas and applications rely on the classical notion of causality that since Bell's theorem \cite{bell1964einstein} we know cannot be applied to quantum phenomena. 

The violation of a Bell inequality shows that quantum correlations are incompatible with the joint assumption of the causal constraints of local realism and measurement independence (``free-will'') \cite{hall2011relaxed,chaves2015unifying}. As it turns out, the phenomenon of quantum nonlocality can be seen as a particular case of causal inference problem \cite{Spekkens2015}, a realization that has sparked a number of generalizations of nonlocality to causal networks of growing size and complexity \cite{fritz2012beyond,chaves2016polynomial,rosset2016,wolfe2019inflation,renou2019genuine}. But apart from the violation of Bell inequalities, are there any other consequences of quantum correlations to the theory of causality?

The standard manner to distinguish between a common cause and direct causal influences among two variables is via an intervention~\cite{pearl2009causal}. However, in some cases it might not be possible to intervene in the system, e.g., due to ethical reasons, or because one is interested in estimating causal effects in past experiments. As shown in Refs.~\cite{ried2015quantum,fitzsimons2015quantum}, differently from the classical case, observed quantum correlations alone are sometimes enough to resolve the question. This has led to a formalization of a quantum common cause \cite{allen2017quantum} and, more generally, quantum causal models \cite{henson2014theory,chaves2015information,costa2016quantum,fritz2016beyond,barrett2019quantum,wolfe2019quantum,aaberg2020semidefinite}. However, the solution in Refs.~\cite{ried2015quantum,fitzsimons2015quantum} relies on causal tomography, that is, it depends on the precise knowledge of the physical system and the measurement apparatuses. Strikingly, as shown in the pioneering work~\cite{balke1997bounds} causal influences can also be estimated without interventions and in a device-independent manner, via the introduction of an instrumental variable. This result, however, relies on the assumption that the unobserved hidden causes are classical and satisfy the property of local realism. In view of that, the instrumental scenario has started to be analyzed from a quantum perspective \cite{chaves2018quantum,van2019quantum,nery2018quantum}, however, despite these initial attempts, it is not known how quantum effects can change the cause and effect relations that can be inferred from the instrumental data. That is precisely the question we resolve in this paper.

We consider the problem of determining casual influences in quantum causal models. To this aim we use the common measure known as the \emph{average causal effect} (ACE) \cite{pearl2009causality}, defined in terms of \emph{interventions}, which can be either measured directly or can be estimated from observational data with the help of an instrumental variable. As we show here, by considering the simplest instrumental scenario, every pure entangled state as well as every pair of incompatible projective measurements can generate correlations that violate the classical bounds on ACE, derived in Ref.~\cite{balke1997bounds}. Remarkably, in this simplest scenario quantum correlations cannot violate any Bell-type inequality \cite{henson2014theory}. That is, our results imply that quantum correlations can generate non-classical signatures going beyond the paradigmatic violation of Bell inequalities. Motivated by that we also introduce a general framework for causal inference in the instrumental scenario, providing bounds for ACE and applicable to quantum theory and beyond.

\emph{Quantifying causality and the instrumental scenario--} Given two variables $A$ and $B$, our aim is to quantify how much of their correlations are due to direct causal influences from $A$ to $B$, or due to some common cause described (classically) by a random variable, $\Lambda$. If we do not have empirical access to the common cause, one option is to intervene on the variable $A$, that is, fix its value to a value of our choice independent of $\Lambda$. The intervention erases any correlation between $A$ and $B$ mediated by $\Lambda$. Thus, any remaining correlation after such intervention can unambiguously be associated to the direct causal influence $A \rightarrow B$. Interventions are a natural choice for quantifying causality. In fact, one of the most widely used measures of causal influence is the ACE measure, defined in terms of interventions as
\begin{equation}
\label{eq:cACE}
    \mathrm{ACE}_{A\rightarrow B} = \max_{a,a',b} \Big(p(b|do(a))-p(b|do(a'))\Big),
\end{equation}
where we used a notation, $p(b|do(a))$ to denote the probability of Bob's outcome $b$ when variable $A$ is set by force to be $a$. We refer to it as \emph{do-probability} in the text. The ACE measures the maximum change in the distribution of the variable $B$ when the value of $A$ is altered. 

\begin{figure}[t!] \centering
		\includegraphics[width=.4\textwidth]{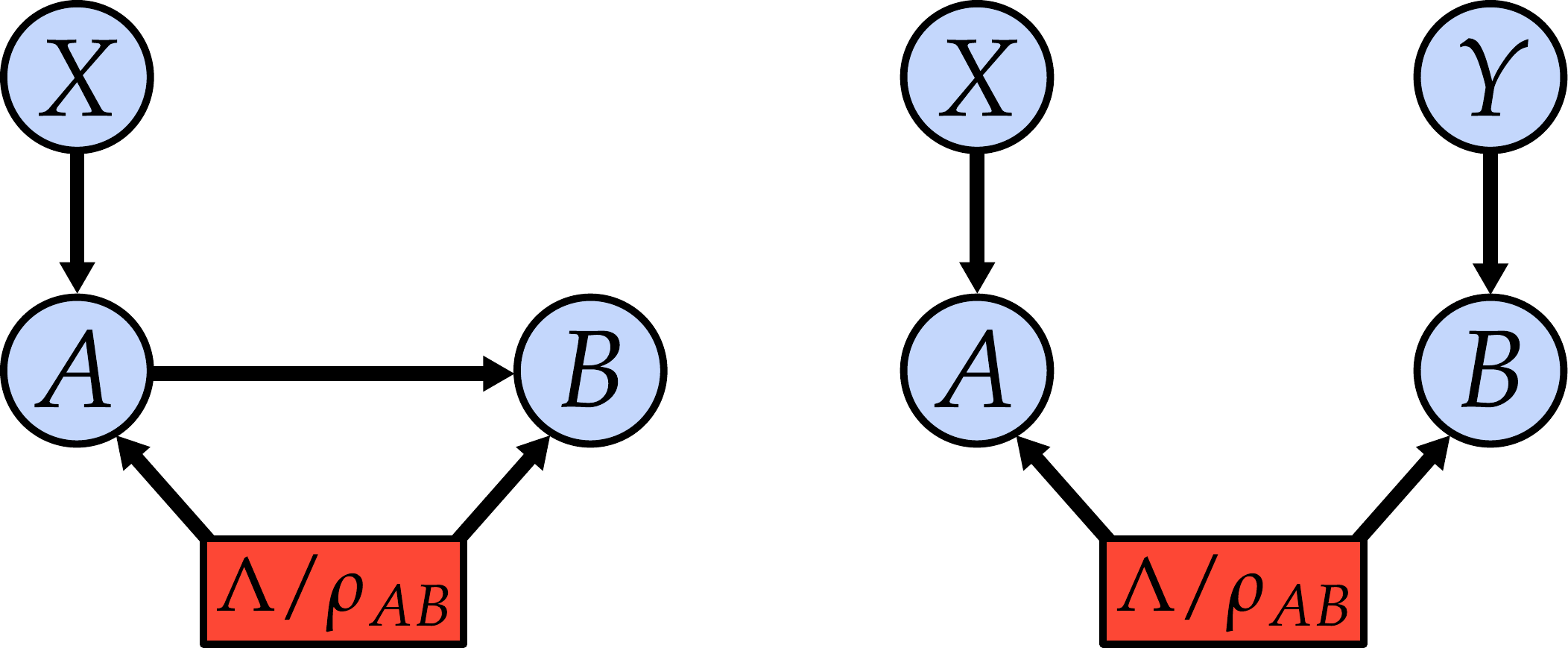}
		\caption{Directed acyclic graphs depicting causal structures:  (left) Instrumental scenario and (right) Bell scenario. In the quantum model we consider here, the classical common source described by a random variable $\Lambda$ is replaced by a quantum (potentially entangled) state $\rho_{AB}$.}
		\label{fig:scenario}
\end{figure}

For a variety of reasons, however, it is not always possible to perform an intervention. With the aim of still being able to estimate causal influences based only on the observational data, the instrumental scenario has been developed \cite{wright1928tariff, angrist1996identification}. The idea is to introduce a third variable in a full control of the experimenter, the so-called instrumental variable $X$. The variable $X$ is assumed to be independent from the common source variable $\Lambda$, that is, $p(x, \lambda) = p(x)p(\lambda)$. Furthermore, $X$ is supposed to have a direct causal effect only over $A$ and not $B$, that is, $p(b|a, x, \lambda) = p(b|a, \lambda)$. Such causal assumptions can be graphically represented via the directed acyclic graph shown in Fig.~\ref{fig:scenario}(left).  It implies that the observed probability distribution is given by
\begin{equation}
\label{eq:classicaldecomp}
p(a,b|x) = \sum_\lambda p(a|x,\lambda)p(b|a,\lambda)p(\lambda).
\end{equation}
Do-probabilities $p(b|do(a))$ are given by
\beq
p(b|do(a)) = \sum_{\lambda}p(b|a,\lambda)p(\lambda),
\eeq
where the conditional distribution $p(b|a,\lambda)$ as well as the distribution of $p(\lambda)$ are the same as in Eq.~(\ref{eq:classicaldecomp}).

To understand the role of the instrumental variable, consider a simple linear relation between the variables given by $b= \kappa a +\lambda$. If we multiply both sides by $x$ and compute the covariance given by $C(X,B)=\avg{X,B}-\avg{X}\avg{B}$, by using $C(X,\Lambda)=0$, we see that $\kappa=C(X,B)/C(A,B)$. That is, simply combining the correlations of $B$ with both $A$ and $X$, we can estimate the causal influence $\kappa$ without the need of any intervention. In this example, however, we assumed a prior knowledge of the functional dependencies among the variables. Nicely, causal influences can be estimated even without such assumptions, just as in the device-independent framework for quantum information \cite{brunner2014bell}, where we perform tasks without the precise knowledge of the underlying physical mechanisms. 

In the particular case where all variables are binary $a,b,x\in\{0,1\}$, the classical ACE (cACE) can be tightly lower-bounded by several expressions including only the observed probabilities $p(a,b\vert x)$ \cite{balke1997bounds}. Here we give one of the bounds that we often use in this work:
\begin{eqnarray}
\label{eq:classbounds}
    & & \mathrm{cACE}_{A\rightarrow B} \geq \\ \nonumber & & 2p(0,0|0)+p(1,1|0)+ p(0,1|1)+p(1,1|1)-2.
\end{eqnarray}
For more lower bounds on $\mathrm{cACE}_{A\rightarrow B}$ see Refs.~\cite{pearl2009causality,balke1997bounds} or Appendix~\ref{app:res1}.

We give another example that signifies the importance of  lower bounds such as in Eq.~(\ref{eq:classbounds}).
Consider that $A$ stands for smoking/non-smoking and $B$ for cancer/no-cancer. Clearly, intervening and forcing people to smoke is not possible. Strikingly, simply introducing an instrumental variable $X$ standing, for example, for $taxation/non-taxation$ of tobacco --that arguably will affect whether people smoke or not, but will not have a direct causal effect on the development of cancer-- and using Eq.~\eqref{eq:classbounds}, we can estimate the effect of interventions and thus lower bound such causal influences.

Within the classical theory of causality, for the bound in Eq.~\eqref{eq:classbounds} to be valid, one needs to assure that the instrumental causal assumptions are fulfilled. In other terms, that the underlying causal structure is that described by Eq.~\eqref{eq:classicaldecomp}. For that aim, the so-called instrumental inequalities have been devised \cite{pearl2009causality,pearl1995testability,bonet2001instrumentality}. 

In the instrumental scenario with binary variables, which we consider here, the only class of instrumental inequalities is given by $ \sum_{a} \max_{x} p(a,b \vert x) \leq~1$~ \cite{pearl1995testability,bonet2001instrumentality}. Curiously, these inequalities remain valid, if the common source is replaced by a quantum state or even post-quantum box~\cite{henson2014theory}, in contrast to the simplest Bell scenario~\cite{clauser1969proposed}.

At first, this might seem to imply that the classical bound on ACE in Eq.~\eqref{eq:classbounds} continues to hold even in the presence of quantum or post-quantum sources. As we show next, this is not the case.

\emph{Quantifying causality with a quantum common source--} If the common source is a bipartite quantum state $\rho_{AB}$, the most general way to generate the classical binary variables $A$ and $B$, is to perform local measurements, described by operators $M_a^x$ and $N_b^a$, on each subsystem. Here the value $x$ is used to choose Alice's measurement setting and the outcome $a$ of Alice's measurement is used to determine Bob's measurement setting, accordingly. Quantum correlations in the instrumental scenario are then described by
\begin{equation}\label{eq:observedcorrelations}
    p(a,b|x)=\Tr[{(M_a^x\otimes N^a_b) \rho_{AB}}].
\end{equation}
In full analogy with the classical case, one can then define quantum interventions as 
\begin{equation}\label{eq:do_prob_quant}
   p(b|do(a))=\Tr{[(\mathbbm{1} \otimes N_b^a) \rho_{AB}]}=\Tr{[N_b^{a} \rho_{B}]},
\end{equation}
where $\rho_B$ is the reduced state of Bob's system. This implies that if an actual intervention is made, the observed quantum average causal effect (qACE) is given by
\begin{equation}\label{eq:qACE}
    \mathrm{qACE}_{A\rightarrow B}=\max_{a,a',b}(\Tr[(N_b^{a}-N_b^{a'})\rho_B]).
\end{equation}
As expected, if the shared state $\rho_{AB}$ is separable, the classical and quantum definitions of ACE coincide (see Appendix~\ref{app:res1}). That is, correlations mediated by a separable state comply with the classical bound in Eq.~\eqref{eq:classbounds}. As stated in our first result, the proof of which can be found in the Appendix~\ref{app:res1}, the same does not hold true for entangled states.

\begin{result}\label{res:ent} Every pure entangled state can generate correlations that violate the classical bound on \emph{ACE}. Moreover, entanglement is necessary but not sufficient for such violations.
\end{result}
This result implies that --even though in the simplest instrumental scenario quantum correlations admit classical explanation of the form in Eq. \eqref{eq:classicaldecomp}-- the amount of observable causal influence $\mathrm{qACE}_{A\rightarrow B}$ is strictly smaller than that required, if the correlations were classical. In other terms, even if no instrumental inequality is violated, the non-classicality of the correlations can be witnessed by interventions on the classical variable $A$. 

\begin{figure}[t!] \centering
		\hspace{-10pt}\includegraphics[width=.5\textwidth]{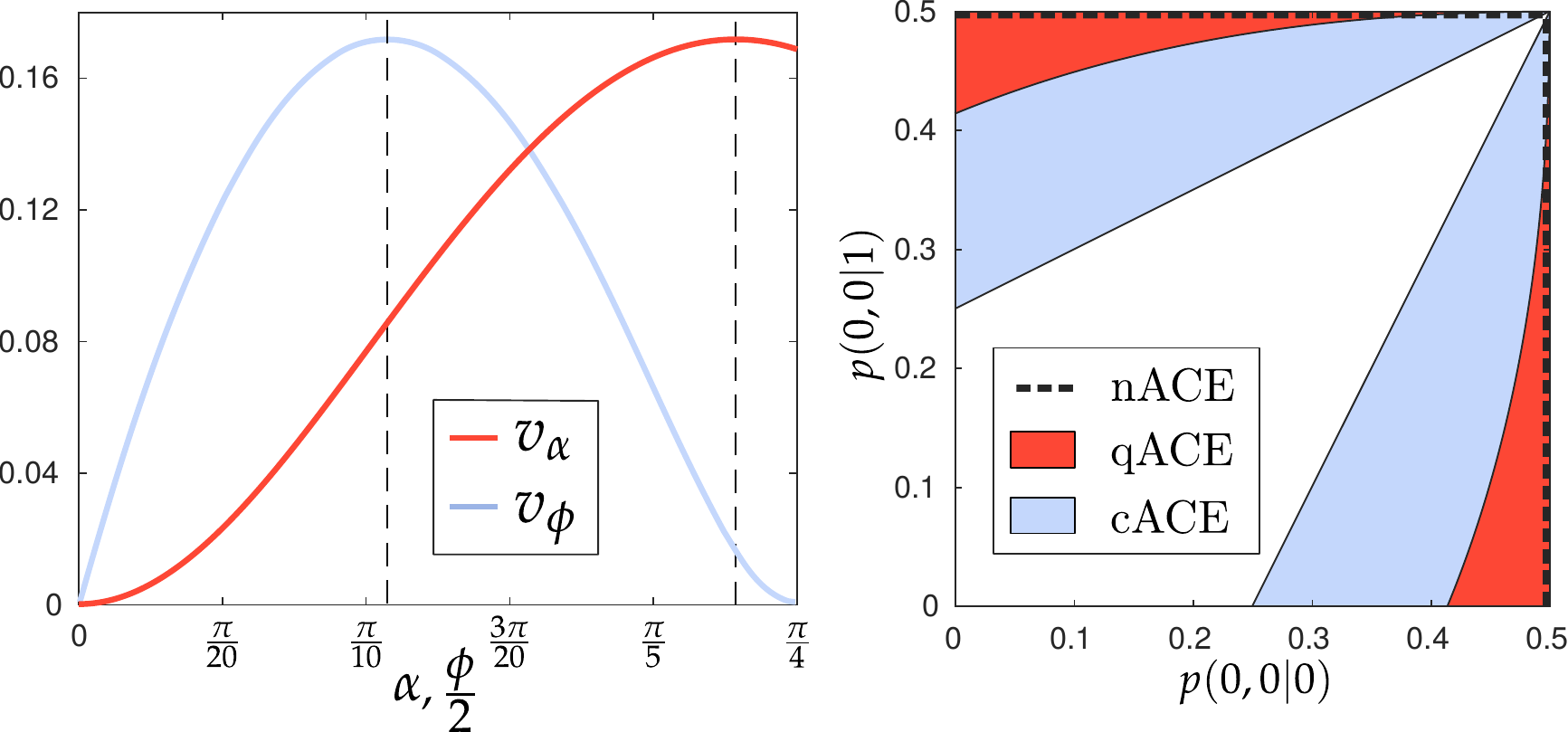}
		\caption{(left) Violation $v_\alpha$ of the classical bound by an entangled two-qubit pure state with parameter $\alpha$; and violation $v_{\phi}$ as a function of the angle $\phi$ between projective measurements of Bob. The dashed lines show that optimal states and measurements are different from the maximal entangled state ($\alpha = \frac{\pi}{4}$) and measurements in mutually unbiased bases ($\phi = \frac{\pi}{4}$). (right) Regions with non-zero lower bounds on cACE (Ref.~\cite{balke1997bounds}), qACE (Eq.~(\ref{eq:qACE_lb})), and nACE (Eq.~(\ref{eq:NSbounds})).}
		\label{fig:regions}
\end{figure}

In order to quantify the degree of violation $v$, we consider how much the classical bound in Eq.~(\ref{eq:classbounds}) overestimates the causal influence in the presence of an entangled source. In Fig.~\ref{fig:regions} we show violation $v_\alpha$ for an entangled two-qubit state $\rho_{AB} = \kb{\psi}{\psi}$, $\ket{\psi} = \cos(\alpha)\ket{0,0}+\sin(\alpha)\ket{1,1}$ for $\alpha\in [0,\frac{\pi}{4}]$.
As detailed in the Appendix~\ref{app:res1}, a maximally entangled two-qubit state violates the classical bound by at most the amount $3(\sqrt{6}-2)/8 \approx 0.169$. However, this is not the optimal violation: non-maximally entangled states give rise to a higher violation up to $3-2\sqrt{2} \approx 0.172$, a fact that in the context of Bell inequalities has been called non-locality anomaly \cite{methot2007anomaly}. Moreover, one can easily see that entanglement is not sufficient for the violation. For example, a maximally entangled state mixed with white noise in the amount of $p$ stays entangled for $p<2/3$, however, it leads to a violation only if $p<1-\sqrt{2/3}\approx 0.1835$.

Violation of Bell inequalities~\cite{clauser1969proposed} is not only a proof that the shared state is entangled, but also a witness of the fact that the measurements being performed should display some non-classicality, as they should be incompatible \cite{khalfin1985quantum,wolf2009measurements,Quintino2014}. As proven in the Appendix~\ref{app:res2} and stated below, a similar result holds for the violation of the classical bounds on causal influence.

\begin{result}\label{res:incomp}
 Every pair of incompatible rank-$1$ projective qubit measurements can generate correlations that violate the classical bound on \emph{ACE}. Moreover, incompatibility of both Alice's and Bob's observables is necessary but not sufficient for the violation. 
\end{result}
In Fig.~\ref{fig:regions} we show violation of the bound in Eq.~(\ref{eq:classbounds}) as a function of the angle $\phi$ between the measurements of Bob that we consider to be $N^a_0=\frac{1}{2}(\openone+\cos(\phi)\sigma_z+(-1)^a\sin(\phi)\sigma_x)$. In Fig.~\ref{fig:regions} the angle $\phi$ ranges between $0$ and $\frac{\pi}{2}$ with $0$ ($\frac{\pi}{2}$) corresponding to perfectly aligned (antialigned) $\sigma_z$ ($\sigma_x$) measurements. The value $\phi=\frac{\pi}{4}$ corresponds to the case of mutually unbiased bases measurements which are optimal for the violation of the simplest Bell inequality~\cite{clauser1969proposed}. In our case, the optimal measurements of Bob correspond to $\phi = \arctan(\frac{2}{\sqrt{3\sqrt{2}+2}})\simeq 0.2149\pi.$

So far we have relied on interventions on the variable $A$ and explicitly taken into account the quantum states and measurements. However, in the more general case we are given some observational data $p(a,b \vert x)$, but do not know a priori which states and measurements have been employed. In this case, our aim is to be able to estimate qACE from the observational data $p(a,b \vert x)$, without actually needing to perform an intervention. That is, in order to find a device-independent bound on qACE, we have to optimize over all possible measurements and states generating the observed correlations $p(a,b \vert x)$. Our approach to this problem is to  map the instrumental scenario to the more familiar and well-studied bipartite Bell scenario \cite{van2019quantum}. 

Let us consider a Bell scenario shown in Fig.~\ref{fig:scenario}(right) that contains the same observed random variables $A,B$ and $X$ as the instrumental scenario in Fig.~\ref{fig:scenario}(left) and an additional classical variable $Y$ that takes values from the same set as $A$, and has a causal effect only on $B$. We also take the hidden common cause, classical or quantum, to be the same for both scenarios. Let $p_\text{Bell}(a,b|x,y)$ be the observed behaviour in the considered Bell scenario. Local hidden-variable theories reproduce correlations of the following type
\begin{equation}
\label{eq:lhv_bell}
p_\text{Bell}(a,b|x,y) = \sum_\lambda p(a|x,\lambda)p(b|y,\lambda)p(\lambda).    
\end{equation}
Conversely, quantum behaviour corresponding to measurement operators $M^x_a$ and $N^y_b$ and quantum state $\rho_{AB}$ is $p_\text{Bell}(a,b|x,y) = \Tr{[(M^x_a\otimes N^y_b) \rho_{AB}]}$. The following mapping
\begin{equation}
\label{eq:map_intr2bell}
p(a,b|x) = p_\text{Bell}(a,b|x,a),\quad \forall a,b,x
\end{equation}
connects classical, quantum and post-quantum correlations in Bell and the instrumental scenarios in a unified manner. Indeed, one can directly see that the mapping in Eq.~(\ref{eq:map_intr2bell}) transforms classical correlations in Eq.~(\ref{eq:lhv_bell}) to the ones in Eq.~(\ref{eq:classicaldecomp}) and the same mapping connects their quantum counterparts. More importantly, we can compute the unobserved do-probabilities $p(b|do(a))$ in terms of $p_\text{Bell}(a,b|x,y)$ in the following way
\begin{equation}
\label{eq:map_do_probs}
p(b|do(a)) = \sum_{a'}p_\text{Bell}(a',b|x,a),\quad \forall a,b,x,
\end{equation}
where the choice of $x$ does not play any role as long as the correlations $p_\text{Bell}(a,b|x,y)$ obey the non-signaling constraints \cite{popescu1994quantum}. One can then see that expressing do-probabilities with the map in Eq.~(\ref{eq:map_do_probs}) is equivalent to the previous definitions for do-probabilities in classical and quantum case. We remark that the mapping in Eqs.~(\ref{eq:map_intr2bell},\ref{eq:map_do_probs}) is not the same as the post-processing on the events of $Y=A$, but is rather a projection from the space of $p_\text{Bell}(a,b|x,y)$ to the space of $p(a,b|x)$ and $p(b|do(a))$. The mapping in Eqs.~(\ref{eq:map_intr2bell},\ref{eq:map_do_probs}) allows the use of known techniques for bounding the set of quantum correlations in Bell scenario, in particular, the so-called NPA hierarchy~\cite{NPA}, with a slight variation: Here the probabilities $p_\text{Bell}(a,b|x,a')$, ($a\neq a')$, with no analogy in the instrumental scenario, play the role of the ``unobserved" variables of the semidefinite program~\cite{vandenberghe1996semidefinite}. Additionally, for binary $A$ one should take into account the relation $p_\text{Bell}(a,b|x,a') = p(b|do(a'))-p(a',b|x)$, that follows from $\sum_a p_\text{Bell}(a,b|x,a') = p_\text{Bell}(b|a')$. 

In the following we focus on the binary case ($a,b,x\in\{0,1\}$) and derive a number of analytical results.

\begin{result}\label{res:qACE_lb}
	In the instrumental scenario with dichotomic measurements \emph{qACE} is lower bounded as
	\begin{eqnarray}\label{eq:qACE_lb}
     \hspace{-0.4cm}\mathrm{qACE}_{A \rightarrow B} \geq &&\sum_{x=0,1}(p(0,0|x)+p(1,1|x))+\zeta-1,
   \end{eqnarray}
   \vspace{-0.5cm}
   \beq
   \zeta = \max_{\pm}\sqrt{\prod_{a=0,1}(1\pm\sum_{x=0,1}(-1)^x(p(a,0|x)-p(a,1|x)))}. \nonumber
    \eeq
\end{result}
The derivation of the above bound is presented in Appendix~\ref{app:res3}. In Fig.~\ref{fig:regions}(right) we compare the lower bounds in Eq.~(\ref{eq:qACE_lb}) and the one in Eq.~(\ref{eq:classbounds}) (along with the other bounds in Ref.~\cite{balke1997bounds}) by plotting the regions in which these bounds are non-zero, showing a clear gap between the classical and quantum descriptions. In Fig.~\ref{fig:regions}(right) a particular slice of the probability space is considered, corresponding to $p(1,0|x)=0,p(0,1|x) = \frac{1}{2}-p(0,0|x), p(1,1|x)=\frac{1}{2}$, $x=0,1$.

\emph{Quantifying causality in post-quantum theories--}
One might be interested whether nontrivial lower bounds similar to Eqs.~(\ref{eq:classbounds},\ref{eq:qACE_lb}) exist in generalized probabilistic theories. Here we answer this question for correlations constrained only by the non-signaling condition in Bell scenario~\cite{popescu1994quantum}.

In order to do so, we map (using Eqs.~(\ref{eq:map_intr2bell},\ref{eq:map_do_probs})) the non-signaling constraints to the instrumental scenario and use linear programming techniques (see Appendix~\ref{app:ns}) to find tight lower bounds on non-signaling ACE  (nACE):   
\begin{equation}\label{eq:NSbounds}
  \hspace{-0.1cm}\mathrm{nACE}_{A\rightarrow B}\geq\max_x(p(0,0|x))+\max_x(p(1,1|x))-1.
\end{equation}
In Fig.~\ref{fig:regions}(right) we also plot the region where $\mathrm{nACE}_{A\rightarrow B}\geq 0$, which is given by two lines with $p(0,0|0) = \frac{1}{2}$ and $p(0,0|1) = \frac{1}{2}$.

\emph{Discussion-- } The incompatibility of quantum correlations with classical causal models is a cornerstone in the foundations of quantum theory. The paradigmatic manner of witnessing this non-classility is via the violation of Bell inequalities. There are causal scenarios, however, where violations of Bell-type inequalities are not possible \cite{henson2014theory}. At first, this might seem to imply that quantum common causes do have a classical explanation in such scenarios. As we show here, this intuition is false. Even in the absence of Bell violations, quantum correlations can violate the classical bounds for the causal influence between two variables in the presence of a quantum common cause. More precisely, every pure entangled state and incompatible projective measurements can violate such bounds. Motivated by this result we propose a general framework to put bounds on the average causal effect in the presence of quantum common causes and even non-signaling boxes. We obtain several analytical results and compare the regions where the aforementioned bounds are non-trivial.

Here we have focused on the scenario where all the observed variables are classical but the common cause can be quantum. Generalizations where other variables in the instrumental causal structure are made quantum open an interesting venue for future research. For instance, the teleportation protocol \cite{bennett1993teleporting} is an instrumental scenario where the instrumental variable $X$ is the state to be teleported and the outcome $B$ is the teleported quantum state. Other paradigmatic quantum information scenarios such as the remote state preparation \cite{bennett2001remote} and dense coding \cite{bennett1992communication} have also an underlying instrumental causal structure. On more foundational side, many physical principles have been developed to understand why quantum correlations do not violate Bell inequalities up to the maximum allowed by special relativity \cite{popescu1994quantum}. In this work we showed that quantum theory also imposes strict bounds on the causal influence between events that differ for generalized probabilistic theories. Can it be that there is an underlying causal principle explaining quantum correlations? We hope that our results will trigger such further developments.

\begin{acknowledgments}
We thank David Gross for fruitful discussions. MG thanks NM for this collaboration and his visit during the pandemic. 
We acknowledge the John Templeton Foundation via the Grant Q-CAUSAL No. 61084, the Serrapilheira Institute (Grant No. Serra-1708-15763), the Brazilian National Council for Scientific and Technological Development (CNPq) via the National Institute for Science and Technology on Quantum Information (INCT-IQ) and Grants No. 307172/2017-1 and No. 406574/2018-9, the Brazilian agencies MCTIC and MEC. We acknowledge partial support by the Foundation for Polish Science (IRAP project, ICTQT, contract no. 2018/MAB/5, co-financed by EU within Smart Growth Operational Programme). MG is funded by the Deutsche Forschungsgemeinschaft (DFG, German Research Foundation) under Germany's Excellence Strategy – Cluster of Excellence Matter and Light for Quantum Computing (ML4Q) EXC 2004/1 – 39053476.
\end{acknowledgments}

\onecolumngrid

\section*{Appendix}
In this Appendix we provide technical details regarding proofs of the results stated in the main text. In Section~\ref{app:res1} we provide a proof of Result~\ref{res:ent} regarding entanglement. In Section~\ref{app:res2} we present a similar result for incompatibility, stated in Result~\ref{res:incomp}. In Section~\ref{app:res3} we explain the derivation of the lower bound on qACE in Eq.~(\ref{eq:qACE_lb}). Finally, we dedicate Section~\ref{app:ns} to a derivation of lower bounds on ACE valid for post-quantum non-signaling theories.

\noindent {\bf Notations:} In the text we denote random variables by capital letters, e.g., $A$, and their values by the corresponding lower-case letters, e.g., $a$. Additionally, we use the common shorthand notation $p(a)\equiv p(A=a)$ for probabilities. We denote POVM effects by capital letters with a double index $M^x_a$, where $x$ stands for the choice of the setting and $a$ for the outcome. Two-outcome POVM is represented by an ordered pair of effects $(M^x_0,M^x_1)$. Binary observables corresponding to this POVM we denote by the same letter with a single index, e.g.,~$M^x = M^x_0-M^x_1$. As in the main text we denote Alice's POVMs by $(M^x_0,M^x_1)$, $x=0,1$ and Bob's by $(N^a_0,N^a_1)$, $a=0,1$.

\subsection{Proof of Result~\ref{res:ent}}\label{app:res1}
We start by restating a result from Ref.~\cite{balke1997bounds}. In the presence of classical common cause the average causal effect $\mathrm{cACE}_{A\rightarrow B}$ can be lower-bounded as follows:
\begin{equation}\label{eq:classbounds_full}
  \mathrm{cACE}_{A\rightarrow B}\ge\max\left\{ \begin{array}{c}
p(0,0|0)+p(1,1|1)-1\\
p(1,1|0)+p(0,0|1)-1\\
2p(0,0|0)+p(1,1|0)+ p(0,1|1)+p(1,1|1)-2\\
p(0,0|0)+2p(1,1|0)+p(0,0|1)+p(1,0|1)-2\\
p(0,1|0)+p(1,1|0)+2p(0,0|1)+p(1,1|1)-2\\
p(0,0|0)+p(1,0|0)+p(0,0|1)+2p(1,1|1)-2
\end{array}\right\} .
\end{equation}
Now we are ready to present proofs of our results on entanglement that we divide into two parts.

\begin{proof}
\textbf{Part 1.} In the first part of the proof  we show that entanglement is necessary to violate the classical bound in Eq.~(\ref{eq:classbounds_full}). The fact that it is not sufficient was discussed in the main text. 

We remind the reader that in the presence of a classical common cause $\Lambda$ we have that
\begin{equation}\label{eq:app_class_decomp}
p(b\vert do(a))= \sum_{\lambda} p(b \vert a,\lambda)p(\lambda),\quad p(a,b \vert x)=  \sum_{\lambda} p(a\vert x,\lambda)p(b \vert a,\lambda)p(\lambda),
\end{equation}
where $p(b \vert a,\lambda)$ and $p(\lambda)$ are the same probability distributions appearing in both decompositions. If the shared state $\rho_{AB}$ is separable, i.e.,~$\rho_{AB}=  \sum_{\lambda} p(\lambda) \rho^{\lambda}_A \otimes \rho^{\lambda}_B$, the observable distribution is given by
\begin{eqnarray}
\label{eq:ent_ness_1}
p(a,b\vert x)= & & ~\tr\left[(M^x_a \otimes N^a_b)\rho_{AB}\right]=\tr\left[(M^x_a \otimes N^a_b) \sum_{\lambda}p(\lambda) \rho^{\lambda}_A \otimes \rho_B^{\lambda}  \right] \\ \nonumber
= & & ~\sum_{\lambda} p(\lambda) \tr\left[M^x_a  \rho^{\lambda}_A\right]\tr\left[N^a_b \rho^{\lambda}_B \right]= \sum_{\lambda} p(a\vert x, \lambda)p(b\vert a, \lambda)p(\lambda),
\end{eqnarray}
where we have identified  $p(a| x,\lambda)=\tr\left[M^x_a  \rho^{\lambda}_A\right]$ and $p(b\vert a, \lambda)=\tr\left[N^a_b\rho^{\lambda}_B \right]$. In turn, an intervention on $A$ leads to
\begin{eqnarray}
p(b\vert do(a))= & & ~\tr\left[ N^a_b \rho_B\right] = \tr\left[ N^a_b \sum_{\lambda} p(\lambda)\rho_B^{\lambda}  \right]= \sum_{\lambda} p(\lambda)\tr\left[ N^a_b  \rho^{\lambda}_B \right] = \sum_{\lambda} p(b\vert a, \lambda)p(\lambda), 
\end{eqnarray}
where again we have $p(b\vert a, \lambda)=\tr\left[ N^a_b \rho^{\lambda}_B\right]$, i.e.,~the same response function as in Eq.~(\ref{eq:ent_ness_1}).

\textbf{Part 2.~} Here we show that every pure entangled state leads to a violation of Eq.~(\ref{eq:classbounds_full}). Let us take $d_A$ and $d_B$ to be the local dimensions of Alice's and Bob's subsystems. Let us then define $D = \min(d_A,d_B)$. We need to consider further two cases, when $D$ is even or odd. The odd case is more involving, but luckily we can consider them together. 
Without loss of generality we consider a bipartite pure state $\rho_{AB} = \kb{\psi}{\psi}$, $\ket{\psi}\in H^{d_A}_A\otimes H^{d_B}_B$, and
\begin{equation}
    \ket{\psi}=\sum^D_{i=1} \lambda_{i} \ket{i,i}, 
\end{equation}
where $\lambda_1\geq \lambda_2 \geq \dots \geq\lambda_D> 0$, and  $\{\ket{i}\}_i$ is the computational basis. Following Ref.~\cite{gisin1992maximal}, we fix the measurements of both parties to be the following:
\beq
M^x = \sin(\theta_x)\bigoplus_{i=1}^{\frac{D}{2}}\sigma_X+\cos(\theta_x)\bigoplus_{i=1}^{\frac{D}{2}}\sigma_Z+\Pi,\quad N^a = \sin(\phi_a)\bigoplus_{i=1}^{\frac{D}{2}}\sigma_X+\cos(\phi_a)\bigoplus_{i=1}^{\frac{D}{2}}\sigma_Z+\Pi,
\eeq
where now we fix the effective dimension of both parties' measurements to be $D$, and $\Pi$ is the matrix with the only non-zero entry $\Pi_{D,D} = 1$, if $D$ is odd and $0$, otherwise. 

Finally, let us define a notation, which turns out to be handy in the subsequent derivations:
\beq
\gamma= \left\{
\begin{array}{l r}
    \lambda_D^2, & D \quad\text{is odd}\\
    0, & \quad\text{otherwise},
\end{array}\right.
\eeq
and
\beq
 M(\theta_x)\equiv M^x = M_0^x-M^x_1,\quad  N(\phi_a)\equiv N^a =N_0^a-N_1^a, \quad x,a\in\{0,1\},
\eeq
Then, we can express qACE as follows:
\begin{align}\label{eq:app_qace}
   \mathrm{qACE}_{A \rightarrow B}= & ~\max_{b}\Big(\Tr[\openone \otimes (N_b^0-N_b^1)\kb{\psi}{\psi}]\Big) = \frac{1}{2}|  \bra{\psi}\openone \otimes (N(\phi_0)-N(\phi_1)\ket{\psi}| \equiv \frac{1}{2}|\mean{\openone \otimes (N(\phi_0)-N(\phi_1)}|.
\end{align}
Without loss of generality we can assume that $\mean{\openone \otimes (N(\phi_0)-N(\phi_1)}\geq 0$. If this is not the case we can consider the scenario where Alice relabels the measurement outcomes and the subsequent derivations would follow accordingly.
Now, let us consider the classical bound in Eq.~(\ref{eq:classbounds}), which we denote as $\mathrm{cACE}^*_{A\rightarrow B}$, (i.e., $\mathrm{cACE}_{A\rightarrow B} \geq \mathrm{cACE}^*_{A\rightarrow B}$),
\begin{align}\label{eq:app_case_bound_d}
     \mathrm{cACE}^*_{A\rightarrow B} = &~ 2 \mean{M^0_0\otimes N^0_0}+\mean{M^0_{1}\otimes N^{1}_{1}}+\mean{M_0^{1}\otimes N^{0}_{1}}+\mean{M_{1}^{1}\otimes N^{1}_{1}}-2\\
     = &~ \frac{1}{2}(1+\mean{\openone\otimes N(\phi_0)}+\mean{M(\theta_0)\otimes \openone}+(1-\gamma)\cos(\theta_0)\cos(\phi_0)+\Lambda \sin(\theta_0)\sin(\phi_0)+\gamma )\nonumber\\
     + &~ \frac{1}{4}(1-\mean{\openone\otimes N(\phi_1)}-\mean{M(\theta_0)\otimes \openone}+(1-\gamma)\cos(\theta_0)\cos(\phi_1)+\Lambda \sin(\theta_0)\sin(\phi_1)+\gamma)\nonumber\\
     + &~ \frac{1}{4}(1-\mean{\openone\otimes N(\phi_{0})}+\mean{M(\theta_1)\otimes \openone}-(1-\gamma)\cos(\theta_1)\cos(\phi_{0})-\Lambda \sin(\theta_1)\sin(\phi_{0})-\gamma)\nonumber\\
    + &~ \frac{1}{4}(1-\mean{\openone\otimes N(\phi_1)}-\mean{M(\theta_1)\otimes \openone}+(1-\gamma)\cos(\theta_1)\cos(\phi_1)+\Lambda \sin(\theta_1)\sin(\phi_1)+\gamma)-2.\nonumber
\end{align}
where $\Lambda=2\sum_{i}\lambda_{2i-1}\lambda_{2i}$. In case of odd $D$ sum in $\Lambda$ truncates at $\lambda_{D-2}\lambda_{D-1}$. For $\mean{M(\theta_x)\otimes N(\phi_a)}$ we used the following identity from Ref.~\cite{gisin1992maximal}:
\begin{equation}
    \mean{M(\theta_x)\otimes N(\phi_a)}=(1-\gamma)\cos(\theta_x)\cos(\phi_a)+\Lambda \sin(\theta_x)\sin(\phi_a)+\gamma.
\end{equation}
We simplify the expression in Eq.~(\ref{eq:app_case_bound_d}) by grouping some of the entries together
\begin{align}\label{eq:app_cace_intermediate}
     \mathrm{cACE}^*_{A\rightarrow B} = & \frac{1}{4}\big(-3+\mean{\openone\otimes N(\phi_0)}+\mean{M(\theta_0)\otimes \openone}-2\mean{\openone\otimes N(\phi_1)}+f(\theta_0,\theta_1,\phi_0,\phi_1,\Lambda,\gamma)\big),
\end{align}
where $f(\theta_0,\theta_1,\phi_0,\phi_1,\Lambda,\gamma)$ is a function which we specify later. Instead, let us first fix $\theta_0=\phi_0$ and use the fact that $N(\phi_0)=M(\phi_0)$, and hence for $\ket{\psi}$, $\mean{\openone\otimes N(\phi_0)}=\mean{M(\phi_0)\otimes \openone}$, to simplify the expression for $\mathrm{cACE}^*_{A\rightarrow B}$ further. 
\begin{align}
     \mathrm{cACE}^*_{A\rightarrow B} = &~ \frac{1}{4}\big(-3+2\mean{\openone\otimes N(\phi_0)} -2\mean{\openone\otimes N(\phi_1)}+f(\theta_1,\phi_0,\phi_1,\Lambda,\gamma)\big)\\
     =&~ \mathrm{qACE}_{A \rightarrow B} +\frac{1}{4}\big(f(\theta_1,\phi_0,\phi_1,\Lambda,\gamma)-3\big),\nonumber
\end{align}
where we inserted the expression for $\mathrm{qACE}_{A\rightarrow B}$ from Eq.~(\ref{eq:app_qace}). If we show that for all entangled states $\ket{\psi}$ we can find angles $\theta_1,\phi_0,\phi_1$ such that $\mathrm{cACE}^*_{A\rightarrow B} > \mathrm{qACE}_{A \rightarrow B}$  we are done. This is equivalent to showing that the function $f(\theta_1,\phi_0,\phi_1,\Lambda,\gamma)-3$ can be made positive for all $\Lambda$ and $\gamma$.

Let us now write $f$ explicitly:
\begin{align}\label{eq:app_f}
    f(\theta_0,\theta_1,\phi_0,\phi_1,\Lambda,\gamma)=2&(1-\gamma)\cos(\theta_0)\cos(\phi_0)+2\Lambda \sin(\theta_0)\sin(\phi_0)+2\gamma\\
    +& (1-\gamma)\cos(\theta_0)\cos(\phi_1)+\Lambda \sin(\theta_0)\sin(\phi_1)+\gamma\nonumber\\
    - & (1-\gamma)\cos(\theta_1)\cos(\phi_{0})-\Lambda \sin(\theta_1)\sin(\phi_{0})-\gamma\nonumber\\
    + &(1-\gamma)\cos(\theta_1)\cos(\phi_1)+\Lambda \sin(\theta_1)\sin(\phi_1)+\gamma.\nonumber
\end{align}
Let us take $\phi_{0}=0$ and $\phi_1=\frac{\pi}{2}$ (and remember that $\theta_0 = \phi_0$). This simplifies the above expression for $f$ and finally we obtain:
\begin{align}\label{eq:app_viol_der_final}
    \mathrm{cACE}^*_{A \rightarrow B}-
     \mathrm{qACE}_{A \rightarrow B}=&~\frac{1}{4}\Big((\gamma-1)(1+\cos{(\theta_1)})+\Lambda\sin{(\theta_1})\Big)\geq \frac{1}{4}\Big(-1-\cos{(\theta_1)}+\Lambda\sin{(\theta_1})\Big)\nonumber\\
     = &~\frac{1}{4}\left(-1+\sqrt{1+\Lambda^2}\bigg(\frac{-1}{\sqrt{1+\Lambda^2}}\cos{(\theta_1)}+\frac{\Lambda}{\sqrt{1+\Lambda^2}}\sin{(\theta_1)}\bigg)\right),
\end{align}
where we used the fact that $\gamma>0$. Thus, if we can prove the statement for even dimensions, for which $\gamma=0$, we are done. 
It is easy to see that in Eq.~(\ref{eq:app_viol_der_final}) the multiplier of $\sqrt{1+\Lambda^2}$ can always be made $1$ by an appropriate choice of $\theta_1$. Finally, we get
\begin{equation}\label{eq:appres1finaleq}
    \mathrm{cACE}^*_{A \rightarrow B}-
     \mathrm{qACE}_{A \rightarrow B}=\frac{1}{4}(\sqrt{1+\Lambda^2}-1)>0.
\end{equation}
The strict inequality holds as long as there are at least two non-zero Schmidt coefficients in $\ket{\psi}$, which is always the case if the state $\ket{\psi}$ is entangled. This completes our proof.

\end{proof}

The violation presented above is not the optimal. Below we consider an arbitrary two-qubit pure entangled state and provide explicit form of measurements, that we argue give the optimal violation of the inequality in Eq.~(\ref{eq:classbounds_full}).
 
Without loss of generality we can fix the state to be $\ket{\psi} = \cos(\alpha)\ket{0,0}+\sin(\alpha)\ket{1,1}$, $\alpha\in [0,\frac{\pi}{4}]$. For this state the reduced density matrices are equal to $\rho_A=\rho_B=\cos^2(\alpha)\kb{0}{0}+\sin^2(\alpha)\kb{1}{1}$. Since we are looking for the optimal violation we go back to the form of the bound as in Eq.~(\ref{eq:app_cace_intermediate}) before we made assumptions about the measurement angles. For two-qubit state this expression takes the form:

\begin{align}\label{eq:appendfinalformof viol}
    \mathrm{cACE}^*_{A \rightarrow B}-
     \mathrm{qACE}_{A \rightarrow B}=&~\frac{1}{4}\big(-3-\mean{\openone\otimes N(\phi_0)}+\mean{M(\theta_0)\otimes \openone}+f(\theta_0,\theta_1,\phi_0,\phi_1,\Lambda,0)\big)\\
     =&~\frac{1}{4}\big(-3+( \cos{\theta_0}-\cos{\phi_0}) \cos(2\alpha)+f(\theta_0,\theta_1,\phi_0,\phi_1,\sin(2\alpha),0)\big),\nonumber
\end{align}
where the function $f(\theta_0,\theta_1,\phi_0,\phi_1,\Lambda,\gamma)$ is given by Eq.~(\ref{eq:app_f}) as before and $\Lambda = \sin(2\alpha)$, $\gamma=0$. Supported by a sufficient numerical evidence we take $\phi_{0}=-\phi_1$ and $\theta_1=-\frac{\pi}{2}$, which seems to be the optimal choice for all $\alpha\in [0,\frac{\pi}{4}]$. The assumption $\phi_{0}=-\phi_1$ also ensures that $\mathrm{qACE}_{A\rightarrow B} = 0$. The function $f(\theta_0,\theta_1,\phi_0,\phi_1,\sin(2\alpha),0)$ now takes the form
\beq
    f(\theta_0,\phi_{0},\sin(2\alpha),0)= 3\cos(\theta_0)\cos(\phi_0)+\sin(2\alpha)\sin(\phi_0)(2+\sin(\theta_0)).
\eeq
Optimally over $\theta_0$ can be resolved analytically and yields
\beq
\theta_{0} = \arccot \left(\frac{\cos (2 \alpha )+3 \cos \left(\phi_0\right)}{\sin (2 \alpha ) \sin \left(\phi_0\right)}\right).
\eeq
The optimal form of $\phi_0$, the only parameter of optimization that is left undetermined, is too unwieldy to be written explicitly here. However, we note that optimization over a single real parameter can be performed up to an arbitrary numerical precision for smooth functions. This optimization leads to the plot of $v_\alpha$ in Fig.~\ref{fig:regions}(left). 

Finally, we also provide the values of $\alpha$ and $\phi_0$
\begin{align}
   \alpha =\arctan\left(\frac{1}{\sqrt{3 \sqrt{2}+2}}\right)+\arctan\left(\sqrt{\frac{1}{2} \left(3 \sqrt{2}+2\right)}\right),\quad \phi_0 =\arctan\left(\frac{2}{\sqrt{3 \sqrt{2}+2}}\right),
\end{align}
which lead to the optimal violation 
\beq
\max_\alpha(v_\alpha) = 3-2\sqrt{2},
\eeq
which we confirm to be the optimal with the help of the NPA hierarchy~\cite{NPA}.

As mentioned in the main text, the maximally entangled state with $\alpha = \frac{\pi}{4}$ does not lead to the optimal violation. Instead, the maximum value that can be attained is equal to $v_{\frac{\pi}{4}} = \frac{3}{8} \left(\sqrt{6}-2\right)$. The proof of this bound can be found in Appendix~\ref{app:res2}. Since $\max_\alpha(v_{\alpha})-v_{\frac{\pi}{4}} \approx 0.00301422$, we conclude that a two qubit Bell state (and the corresponding optimal measurements) does not lead to the maximal violation of the classical bound on ACE.

\subsection{Proof of Result~\ref{res:incomp}}
\label{app:res2}
Below we give a proof of Result~\ref{res:incomp} regarding incompatibility from the main text. 
For convenience, we provide the definition of compatibility of two-outcome POVMs $\{(M^x_0,M^x_1)\}_x$ (see e.g., Ref.~\cite{ali2009commutative}).
\begin{definition}[POVM compatibility]
A collection of two-outcome POVMs $\{(M^x_0,M^x_1)\}_x$ is said to be compatible (or jointly measurable) if there exist a so-called parent POVM $(G_0,\dots,G_{m_\lambda})$, ($G_\lambda\geq 0$, $\sum_\lambda^{m_\lambda} G_\lambda = \openone$), and a classical post-processing function $d(a|x,\lambda)$, ($d(a|x,\lambda)\geq 0$, $\sum_{a=0,1} d(a|x,\lambda)=1,\; \forall x,\lambda$), such that the following holds
\beq
\label{eq:incomp_def}
M^x_a = \sum_\lambda^{m_\lambda} d(a|x,\lambda) G_\lambda,\quad \forall a,x.
\eeq
\end{definition}
The proof is divided into two parts. 

\begin{proof}\textbf{Part 1.} In the first part we show that compatibility of Alice's or Bob's measurements would necessarily lead to compliance of ACE with the classical bound in Eq.~(\ref{eq:classbounds_full}). 
Let us first prove that if Alice's POVMs $(M^0_0,M^0_1)$,$(M^1_0,M^1_1)$ admit the decomposition in Eq.~(\ref{eq:incomp_def}), no violation of classical ACE bound can be observed. Let $M^x_a = \sum_\lambda^{m_\lambda} d(a|x,\lambda) G_\lambda$, $a,x\in \{0,1\}$, then the observed quantum behaviour $p(a,b|x)$ takes the form
\beq
p(a,b|x) = \sum_\lambda d(a|x,\lambda)\tr[(G_\lambda\otimes N^a_b)\rho_{AB}] = \sum_\lambda d(a|x,\lambda)\tr[\sigma_\lambda N^a_b] = \sum_\lambda d(a|x,\lambda)\tr[\hat{\sigma}_\lambda N^a_b]\tr(\sigma_\lambda),
\eeq
where we have introduced the notation $\sigma_\lambda = \tr_{A}[(G_\lambda\otimes\openone)\rho_{AB}]$ and $\hat{\sigma}_\lambda = \frac{\sigma_\lambda}{\tr[\sigma_\lambda]}$. Since $\sum_\lambda G_\lambda = \openone$, we know that $\sum_\lambda\tr[\sigma_\lambda]=1$, i.e. $\{\tr[\sigma_\lambda]\}_\lambda$ defines a probability distribution. On the other hand, since each $\hat{\sigma}_\lambda$ is a normalized state, $\sum_b\tr[\hat{\sigma}_\lambda N^a_b] = 1, \forall a,\lambda$. Hence, we constructed a decomposition of the behaviour $p(a,b|x)$ in the form of Eq.~(\ref{eq:classicaldecomp}). At the same time, the do-probability $p(b|do(a))$ can be decomposed as follows:
\beq
p(b|do(a)) = \tr[(\openone\otimes N^a_b)\rho_{AB}] = \sum_\lambda\tr[(G_\lambda\otimes N^a_b)\rho_{AB}] = \sum_\lambda\tr[\hat{\sigma}_\lambda N^a_b]\tr[\sigma_\lambda],
\eeq
i.e., it admits the decomposition in Eq.~(\ref{eq:app_class_decomp}) with the same response function $\tr[\hat{\sigma}_\lambda N^a_b]$ and distribution $\{\tr[\sigma_\lambda]\}_\lambda$ of $\lambda$ as the behaviour $p(a,b|x)$. From here, it follows that the classical bound in Eq.~(\ref{eq:classbounds_full}) holds for ACE for compatible measurements of Alice. 

Similarly, we can repeat the above construction for the case when Bob's POVMs are jointly measurable, i.e., when $N^a_b$ admit a decomposition of the form $N^a_b = \sum_\lambda^{m_\lambda} d(b|a,\lambda) G_\lambda,\;\forall a,b$. Let us again write the behaviour $p(a,b|x)$
\beq\label{eq:app_incmop_bob}
p(a,b|x) = \sum_\lambda d(b|a,\lambda)\tr[(M^x_a\otimes G_\lambda) \rho_{AB}] = \sum_\lambda d(b|a,\lambda)\tr[M^x_a\hat{\sigma}_\lambda]\tr[\sigma_\lambda],
\eeq
where now we denoted $\sigma_\lambda = \tr_B[(\openone\otimes G_\lambda)\rho_{AB}]$, and again $\hat{\sigma}_\lambda = \frac{\sigma_\lambda}{\tr[\sigma_\lambda]}$. We have already shown above that $\{\tr[\sigma_\lambda]\}_\lambda$ is a valid probability distribution and again due to normalization of $\hat{\sigma}_\lambda$ we conclude that $\tr[M^x_a\hat{\sigma}_\lambda]$ is a valid response function. For the do-probabilities, we can directly conclude that 
\beq
p(b|do(a)) = \sum_\lambda d(b|a,\lambda) \tr[(\openone\otimes G_\lambda)\rho_{AB}] = \sum_\lambda d(b|a,\lambda) \tr[\sigma_\lambda],
\eeq
i.e., they satisfy the decomposition in Eq.~(\ref{eq:app_class_decomp}) with the same response functions as in Eq.~(\ref{eq:app_incmop_bob}).

\textbf{Part 2.} In the second part we show that for an appropriate choice of a quantum state $\rho_{AB}$ and measurements of one party, the incompatibility of projective measurement of the other party can be witnessed by the violation of Eq.~(\ref{eq:classbounds}). We would like to rewrite the inequality in terms of the effects corresponding to the ``$0$" outcome only. This is, of course, possible due to the normalization: $M^x_1=\openone-M^x_0$ and $N^a_1=\openone-N^a_0$. As a result, we obtain the following inequality:
\beq\label{eq:app_ace_ineq_incomp_1}
\tr\big[\rho_{AB}(-\openone\otimes(N^0_0+N^1_0)-M^0_0\otimes\openone + 2M^0_0\otimes N^0_0+M^0_0\otimes N^1_0-M^1_0\otimes N^0_0+M^1_0\otimes N^1_0)\big]\leq 0,
\eeq
Now let us assume that the state $\rho_{AB}$ is the maximally entangled state, i.e., $\rho_{AB} = \frac{1}{2}(\kb{00}{00}+\kb{00}{11}+\kb{11}{00}+\kb{11}{11})$, and $\tr[M^x_a]=1, \tr[N^a_b] = 1, \forall a,b,x$. This transforms the inequality in Eq.~(\ref{eq:app_ace_ineq_incomp_1}) to the following form
\beq
-3 + \tr\big[(M^0_0)^T(2N^0_0+N^1_0)\big]+\tr\big[(M^1_0)^T(N^1_0-N^0_0)\big]\leq 0,
\eeq
where $(\cdot)^T$ stands for a transposition with respect to the basis in which $\rho_{AB}$ was defined. We can now take $(M^0_0)^T$ and $(M^1_0)^T$ to be the eigenstates of the operators $2N^0_0+N^1_0$ and $N^1_0-N^0_0$, respectively, that correspond to the largest eigenvalues of these operators. Finally, for qubit POVMs we can write Bloch decomposition of the effects $N^a_0 = \frac{\openone}{2}+\frac{\vec{n}_a\cdot\vec{\sigma}}{2}$, which leads to the following inequality for Bloch vectors of Bob's measurements
\beq\label{eq:app_incomp_nn}
\frac{1}{4}\left(|2\vec{n}_0+\vec{n}_1|+|\vec{n}_0-\vec{n}_1|-3\right)\leq 0.
\eeq

One can see that the above inequality is violated by all incompatible rank-1 PVMs $(N^0_0,N^0_1)$ and $(N^1_0,N^1_1)$, which is the case whenever $\vec{n}_0\cdot\vec{n}_1\neq \pm 1$. Moreover, for noisy POVMs with $|\vec{n}_0| = |\vec{n}_1|$, the above inequality can still be violated by some POVMs whenever $|\vec{n}_0|\geq \sqrt{\frac{2}{3}}$. It is easy to see that the same analysis can be carried out for Alice's measurements.
\end{proof}
We can continue the above calculations to derive the upper-bound on the violation of inequality in Eq.~(\ref{eq:classbounds}) by a maximally entangled state. So far we made only one assumption that $\tr[M^x_a]=1, \tr[N^a_b] = 1, \forall a,b,x$, which is always the case for rank-$1$ projective measurements, which are the extremal two-outcome qubit measurements. The norms in Eq.~(\ref{eq:app_incomp_nn}) can be easily calculated for normalized vectors $\vec{n_0},\vec{n_1}$, which leads to the following expression for the violation 
\beq
\frac{1}{4}\left(\sqrt{5+4\vec{n}_0\cdot\vec{n}_1} + \sqrt{2-2\vec{n}_0\cdot\vec{n}_1}-3\right).
\eeq
A simple maximization over the inner-product $\vec{n}_0\cdot\vec{n}_1$ gives the value $v_{\frac{\pi}{4}}$ stated in Appendix~\ref{app:res1}.

\subsection{Proof of 
Result~\ref{res:qACE_lb}}\label{app:res3}
In this section of the Appendix we derive the  lower bound on quantum average causal effect in Eq.~(\ref{eq:qACE_lb}). We start by writing a general Bell expression for two dichotomic measurements:
\beq\label{eq:bell_arb}
\mathcal{B}(\alpha,\beta,\gamma,\delta) = -\alpha\langle M^0\otimes N^0\rangle+\beta\langle M^0\otimes N^1\rangle+\gamma\langle M^1\otimes N^0\rangle + \delta\langle M^1\otimes N^1\rangle, 
\eeq
where $M^x = M^x_0-M^x_1$, $N^y = N^y_0-N^y_1$ are Alice's and Bob's observables and $\langle M^x\otimes N^y\rangle = \tr[(M^x\otimes N^y)\rho_{AB}]$. Real coefficients $\alpha,\beta,\gamma$ and $\delta$ are at the moment not specified. Since the dimensions of observables and the state are not restricted, we can assume $\rho_{AB}$ to be pure and the measurements to be projective. Then one can upper-bound the expression in Eq.~(\ref{eq:bell_arb}) in the following way: First, we group the terms corresponding to observables $M^0$ and $M^1$ and afterwards, we use the Cauchy–Schwarz inequality to obtain the following:
\beq
\mathcal{B}(\alpha,\beta,\gamma,\delta)\leq \sqrt{\langle (M^0)^2\otimes\openone\rangle}\sqrt{\langle \openone\otimes(\alpha N^0-\beta N^1)^2\rangle}+\sqrt{\langle (M^1)^2\otimes\openone\rangle}\sqrt{\langle \openone\otimes(\gamma N^0+\delta N^1)^2\rangle}.
\eeq
Since $M^x$ and $N^y$ are two-outcome projective observables, we have that $(M^x)^2=(N^y)^2=\openone$ and we can rewrite the inequality as follows:
\beq\label{eq:app_abgd_1}
\mathcal{B}(\alpha,\beta,\gamma,\delta)\leq \sqrt{\alpha^2+\beta^2-\alpha\beta\tr[\rho_{AB}\openone\otimes (N^0N^1+N^1N^0)]}+\sqrt{\gamma^2+\delta^2+\gamma\delta\tr[\rho_{AB}\openone\otimes (N^0N^1+N^1N^0)]}.
\eeq

On the other hand, we can express $\mathcal{B}(\alpha,\beta,\gamma,\delta)$ in terms of the observed correlations $p(a,b|x)$ and the do-probabilities $p(b|do(a))$ as follows
\beq\label{eq:app_abgd_1a}
\mathcal{B}(\alpha,\beta,\gamma,\delta) =&& -\alpha-\beta+\gamma-\delta+2p(0|do(0))(\alpha-\gamma)+2p(0|do(1))(\beta+\delta)-2\alpha(p(0,0|0)-p(0,1|0))\nonumber\\
&&+2\beta(p(1,1|0)-p(1,0|0))+2\gamma(p(0,0|1)-p(0,1|1))+2\delta(p(1,1|1)-p(1,0|1)).
\eeq
Notice that for  $\gamma=1+\alpha$ and $\delta=1-\beta$, in Eq.~(\ref{eq:app_abgd_1a}) the expression of do-probabilities form the qACE. Hence, setting these values for $\gamma$ and $\delta$, we can express qACE from Eq.~(\ref{eq:app_abgd_1a}) as a function of $\alpha,\  \beta$, correlations,  $p(a,b|x)$ and the value $\mathcal{B}(\alpha,\beta,1+\alpha,1-\beta)$. 

Next we derive an upper bound in Eq.~(\ref{eq:app_abgd_1}). In particular, we  maximize the right-hand side of Eq.~(\ref{eq:app_abgd_1}) with respect to the real parameter $\xi\equiv \tr[\rho_{AB}\openone\otimes (N^0N^1+N^1N^0)]$, subject to the following constraints   $-2\leq \xi\leq 2$ and find that  the optimal solution corresponds to the following expression of $\xi$:
\beq\label{eq:app_abgd_xi}
\xi = \frac{\alpha}{\beta}+\frac{\beta}{\alpha}+\frac{\alpha+\beta}{1+\alpha}+\frac{\alpha+\beta}{\beta-1}.
\eeq
Inserting this expression in the inequality leads to the bound
\beq\label{eq:app_abgd_2}
\mathcal{B}(\alpha,\beta,1+\alpha,1-\beta) \leq |\alpha+\beta|\left(\sqrt{\frac{(1+\alpha)(1-\beta)}{\alpha\beta}}+ \sqrt{\frac{\alpha\beta}{(1+\alpha)(1-\beta)}}\right),
\eeq
which is applicable when $-2\leq\xi\leq 2$.  Combining Eq.~(\ref{eq:app_abgd_1}) (for $\gamma=1+\alpha$ and $\delta=1-\beta$) and Eq.~(\ref{eq:app_abgd_2}) allows to obtain the following lower bound on qACE:
\beq\label{eq:app_abgd_3}
\mathrm{qACE}_{A\rightarrow B}\geq&& \,2(p(0,0|1)+p(1,1|1))-1-\alpha\sum_{x=0,1}(-1)^x(p(0,0|x)-p(0,1|x))-\beta\sum_{x=0,1}(-1)^x(p(1,0|x)-p(1,1|x))\nonumber\\
&&-\frac{|\alpha+\beta|}{2}\left(\sqrt{\frac{(1+\alpha)(1-\beta)}{\alpha\beta}}+ \sqrt{\frac{\alpha\beta}{(1+\alpha)(1-\beta)}}\right),
\eeq
where $\alpha$ and $\beta$ need to satisfy the condition that  $\-2\leq \xi\leq 2$ and any such assignment of $\alpha$ and $\beta$  leads to a valid lower bound on qACE. Naturally, in order to obtain a tighter bound, we try to maximize the right-hand side of Eq.~(\ref{eq:app_abgd_3}) with respect to $\alpha$ and $\beta$. This, however, turns out to be not a straightforward task, given that the correlations $p(a,b|x)$ are to remain unspecified. Here we consider a specific fruitful assignment that complies with $2\leq \xi \leq 2$ for all $\alpha$
\beq
\beta = \frac{1+\alpha}{1+2\alpha}.
\eeq
Making the above substitution leaves us with an expression of the lower bound carrying merely an unconstrained  parameter, $\alpha$ and the observed probabilities.  Such an optimization over $\alpha$ is then possible and results in two solutions (depending on the choice of $\pm$):
\beq\label{eq:app_abgd_4}
\alpha = \frac{1}{2}\left(\pm\sqrt{\frac{\pm 1 + p(1,0|0) - p(1,0|1) - p(1,1|0) + p(1,1|1)}{\pm 1 + p(0,0|0) - p(0,0|1) - p(0,1|0) + p(0,1|1)}}-1\right).
\eeq
One should note here that,  due to the constraints on the observed correlations in the instrumental scenario, namely, $p(1,1|0)+p(1,0|1)\leq 1$ and $p(0,1|0)+p(0,0|1)\leq 1$, the expression under the square root in Eq.~(\ref{eq:app_abgd_4}) is always non-negative. Finally, by substituting the expression of $\alpha$ in Eq.~(\ref{eq:app_abgd_3}) we directly obtain the expression in Result~\ref{res:qACE_lb}.

\subsection{Bounds on ACE for non-signaling probabilistic theories}\label{app:ns}
Here we explain the derivations of the lower-bound in Eq.~(\ref{eq:NSbounds}) on nACE. Without loss of generality let us assume that $p(0|do(0))\geq p(0|do(1))$. From a simple relation between marginal probabilities in Bell scenario, namely, $p_\text{Bell}(b|y) \geq p_\text{Bell}(a,b|x,y)$, we conclude that 
\beq
p(b|do(a)) \geq \max_x(p(a,b|x)),\; \forall a,b,
\eeq
and in particular $p(0|do(0))\geq \max_x(p(0,0|x))$. It also follows that
\begin{equation}\label{eq:app_ns_ineq_1}
p(0|do(1))=1-p(1|do(1))\leq 1-\max_x p(1,1|x).   
\end{equation}
Combining both bounds above leads to the bound given in Eq.~(\ref{eq:NSbounds}):
\begin{align}\label{eq:app_ns_ineq_2}
\mathrm{nACE}_{A\rightarrow B} & \geq \max_x(p(0,0|x)) +\max_x (p(1,1|x))-1.
\end{align}

Alternatively, one could use linear programming techniques in order to reproduce the above result. Namely, one could use convex polytope software (e.g., Panda \cite{lorwald2015panda}) in order to obtain the extremal points of non-signaling polytope in Bell scenario. Then one would apply the mapping in Eqs.~(\ref{eq:map_intr2bell},\ref{eq:map_do_probs}), which is an affine transformation in the space of probability vectors. Using the same software one more time, one could obtain the inequalities that determine the relations between the probabilities $p(a,b|x)$ and do-probabilities $p(b|do(a))$ in the instrumental scenario with non-signaling source of correlations. These inequalities turn out to be exactly the ones in Eqs.~(\ref{eq:app_ns_ineq_1},\ref{eq:app_ns_ineq_2}). The resulting bound is again given by Eq.~(\ref{eq:NSbounds}), which also proves its tightness.

\twocolumngrid

\bibliography{bibliography}

\end{document}